\documentclass[twocolumn,showpacs,preprintnumbers,amsmath,amssymb]{revtex4}

\usepackage{graphicx}
\usepackage{dcolumn}
\usepackage{bm}

\def\bea{\begin{eqnarray}}
\def\eea{\end{eqnarray}}

\begin{document} 

\preprint{Version 1.9}


\title{The RHIC azimuth quadrupole: ``perfect liquid'' or gluonic radiation?}

\author{Thomas A. Trainor}
\address{CENPA 354290, University of Washington, Seattle, WA 98195}


\date{\today}

\begin{abstract}
Large elliptic flow at RHIC seems to indicate that ideal hydrodynamics provides a good description of Au-Au collisions, at least at the maximum RHIC energy. The medium formed has been interpreted as a nearly perfect (low-viscosity) liquid, and connections have been made to gravitation through string theory. Recently, claimed observations of large flow fluctuations comparable to participant eccentricity fluctuations seem to confirm the ideal hydro scenario. However, determination of the azimuth quadrupole with 2D angular autocorrelations, which accurately distinguish ``flow'' (quadrupole) from ``nonflow'' (minijets), contradicts conventional interpretations. Centrality trends may depend only on the initial parton geometry, and methods used to isolate flow fluctuations are sensitive instead mainly to minijet correlations. The results presented in this paper suggest that the azimuth quadrupole may be a manifestation of gluonic multipole radiation.
\end{abstract}


\maketitle

 \section{Introduction}

The large elliptic flow $v_2$ observed in RHIC heavy ion collisions compared to lower energies and to hydro predictions is said to reveal a ``perfect liquid''~\cite{perfliq,rom}. Elliptic flow is interpreted as a hydrodynamic response to early pressure and the azimuthal eccentricity $\epsilon$ of the initial system~\cite{ollitrault}. In the hydrodynamic (hydro) model of nuclear collisions large flow values imply small viscosity and rapid thermalization~\cite{hydro}. However, the applicability of a hydrodynamical model to nuclear collisions and the true nature of the azimuth quadrupole interpreted as elliptic flow can be questioned. In this paper I examine analysis techniques and interpretations leading to ``perfect liquid'' and describe some alternatives.

\subsection{Conventional flow argument}

The conventional flow argument is summarized as follows~\cite{ollitrault,teaney}: Azimuthal asymmetry of participant nucleons at initial nuclear contact,  copious particle production, subsequent thermalization by rescattering, and hydrodynamic evolution lead to a matching asymmetry in final-state momentum space, providing direct evidence for strong interactions among initially-produced particles (gluons). Thermalization must be rapid because hydro expansion reduces the initial space asymmetry. The argued connection between initial and final states through rapid thermalization is thus demonstrated by large elliptic flow and the apparent success of the hydro model, especially the relation $v_2 \propto \epsilon$ for ideal hydro.

It follows that participant-geometry (eccentricity) fluctuations and flow fluctuations should manifest the same $v_2 \propto \epsilon$ relationship. Recent participant-based eccentricity models seem to unify flow measurements across collision systems~\cite{manlyeps}, and flow fluctuation measurements seem to provide an exact connection between flow measure $v_2$ and {\em participant} eccentricity $\epsilon_{part}$~\cite{starflucts,phobflucts}. However, modeling low-$x$ partons with a participant-nucleon eccentricity is questionable at mid-rapidity, and the true magnitude of $v_2$ fluctuations remains unclear. 

\subsection{Open issues for flow analysis}

Observation of the trend $v_2 \propto \epsilon$ would demonstrate that HI collisions at RHIC produce a thermalized QCD medium (quark-gluon plasma or variant) according to ideal hydro~\cite{ollitrault,hydro}. Recent experimental results suggest that $v_2 \propto \epsilon$ has been achieved~\cite{cipanp}. However, there remain open issues: 1) Conventional flow measure $v_2$, motivated by the hydro scenario~\cite{hydro}, admits significant statistical bias (systematic error in numerical procedures), especially for smaller event multiplicities, and is difficult to interpret~\cite{flowmeth}. 2) Conventional flow measurements do not distinguish reliably between flow and ``nonflow,'' identified as angular correlations from low-$Q^2$ parton (gluon) fragmentation or {\em minijets}~\cite{ppcorr, axialci,ptscale,edep,lepmini}. 3) flow fluctuation measurements have been based on assumptions about ``nonflow''  known to be invalid from minijet studies.

4)  Recent A-A eccentricity definitions are based on simulated distributions of point-like participant nucleons. Modeling the transverse distribution of low-$x$ partons at RHIC midrapidity with point-like nucleons is questionable. Are assumptions and results consistent with known parton distributions, with the concept of a color-glass condensate? 5) Conventional flow measurements and terminology are strongly model dependent. Imposition of an {\em a priori} hydro model on data may obscure more fundamental processes.

\subsection{Possible resolutions}

Two new initiatives are needed. First, alternative physical mechanisms for azimuth multipoles should be explored. We should consider the analog in QCD to multipole radiation fields in electromagnetism, especially given the non-Abelian nature of the gluonic field. What is the long-wavelength limit of QCD, the equivalent of Maxwell's equations? A static-field limit in the form of a color-glass condensate (CGC) has been introduced~\cite{cgc}, but its dynamical implications have not been fully explored. Gluonic multipole fields may play a significant role in nuclear collisions and may produce strong azimuth correlations.

Second, new correlation analysis methods include a model-independent statistical measure and analysis procedure which reliably distinguish the azimuth quadrupole moment (model-independent terminology) attributed to elliptic flow from ``nonflow''~\cite{flowmeth}. Accurate measurements  of the azimuth quadrupole over a range of centralities, energies and nuclear sizes may provide new insights into the phenomenon described as elliptic flow.

\subsection{Paper outline}

In this paper I review conventional and 2D autocorrelation analysis methods applied to measurement of elliptic flow or the azimuth quadrupole component, including power spectra, Pearson's normalized covariance and improved A-A centrality methods~\cite{flowmeth}. I review two approaches to A-A eccentricity simulation and advocate the optical Glauber model. I compare published STAR $v_2$ data and fits to simulated 2D angular autocorrelations. I conclude that what is termed ``nonflow'' is dominated by the same-side minijet peak (jet cone) and that there may be a simple  relationship between the true azimuth quadrupole moment and initial-state geometry parameters. I show that recently claimed large flow fluctuations are probably dominated by minijets. Finally, I review problematic issues for hydrodynamic models and suggest that the azimuth quadrupole is actually a manifestation of QCD field-field interactions, possibly related to longitudinal filimentation and instabilities, that elliptic flow is actually gluonic multipole radiation.


\section{Analysis Methods}

Issues for azimuth correlation analysis include the correlation measure,  the projection method to angular subspaces,  the application of Fourier series,  centrality measurement and plotting formats. Correlation analysis methods are also discussed in~\cite{florence,inverse}.

\subsection{Conventional method {\em vs} alternatives}

Conventional flow analysis is defined in terms of single-particle density $\rho$ on azimuth angle $\phi$ relative to reaction-plane angle $\Psi_r$ estimated by {\em event plane} angle $\Psi_m$. Flow measure $v_2$ associated with the $m=2$ Fourier term is obtained by several methods (e.g., event plane, subevents, two-particle correlations, four-particle cumulants)~\cite{flowmeth}. Because of a factorization in the definition of the Fourier series~\cite{poskvol}  $v_2$ is actually the square root of the {\em ratio} of two pair densities, a {\em per-pair} two-particle correlation measure~\cite{flowmeth} with the form $v_2 = \sqrt{\Delta \rho[2] / \rho_{ref}}$, where $\Delta \rho  = \rho - \rho_{ref}$ is the density of correlated pairs, $\rho_{ref}$ is the reference (mixed) pair density, and `[2]' denotes the ``second harmonic'' or second Fourier coefficient of $\Delta \rho$. The result is a two-particle azimuth correlation analysis directed to a specific sinusoid amplitude.

The {\em event-plane} method uses event-wise flow vector $\vec{Q}_2$ to estimate the reaction plane by event-plane angle $\Psi_2$. Intermediate value $v_{2,observed} = \langle \cos(2[\phi - \Psi_2])\rangle$ is then corrected by an estimate of the event-plane resolution. The EP procedure is closely related to a standard 1D azimuth autocorrelation analysis measuring the equivalent of $v_2\{2\}$. Differences between $v_2\{EP\}$ and $v_2\{2\}$ arise because the EP procedure is an approximation to the 1D autocorrelation~\cite{flowmeth}. 

The 2D quadrupole moment analysis is based on the power spectrum derived from the {azimuth autocorrelation} by a Fourier transform according to the Wiener-Khintchine theorem~\cite{flowmeth}. When extended to 2D angular autocorrelations on difference axes $(\eta_\Delta,\phi_\Delta)$ the autocorrelation analysis is able to distinguish ``elliptic flow'' accurately from so-called ``non-flow'' dominated by minijets~\cite{minijets}. The four-particle cumulant $v_2\{4\}$ developed to suppress nonflow~\cite{borg} approximates results from a 2D autocorrelation analysis.

\subsection{Correlation measures and Pearson's covariance}

The {\em per-particle} measure of azimuth correlations $\Delta \rho[2] / \sqrt{\rho_{ref}}$ has the form of Pearson's correlation coefficient or normalized covariance converted to a density. The `[2]' denotes the {\em quadrupole} component of azimuth correlations. Its relation to $v_2$ is defined below and in~\cite{flowmeth}. The {\em interpretability} of per-particle correlation measures, their success in revealing the details of nuclear collision dynamics, is described in~\cite{axialci,axialcd,ptscale,edep}.

Pearson's covariance is invariant under {\em linear superposition}. Its value for a combination of {\em independent and equivalent} systems is the same as that for  any component. Thus, changes in the measure under composition (e.g., A-A centrality variation) indicate {real} physical changes in the composite and truly differential correlation measurement. Because $v_2$ contains an extraneous factor $1/\sqrt{n_{ch}}$, variations with energy and centrality can be misleading (e.g., cf.~\cite{edep}).

\subsection{2D angular autocorrelations}

Two-particle densities are defined on 6D momentum space $(p_{t1},\eta_1,\phi_1,p_{t2},\eta_2,\phi_2)$. In this paper I restrict to $p_t$-integrated pair distributions on 4D angular subspace $(\eta_1,\phi_1,\eta_2,\phi_2)$. Further projection of the 4D space to 2D subspace $(\eta_\Delta,\phi_\Delta)$ (e.g. $\eta_\Delta = \eta_1 - \eta_2$, $\eta_\Sigma = \eta_1 + \eta_2$) is described in~\cite{inverse,flowmeth}. Within a limited rapidity interval near mid-rapidity {\em stationarity} is usually valid: the correlation structure is approximately uniform on pair mean pseudorapidity $\eta_\Sigma/2$, and we expect stationarity to hold on azimuth for a $2\pi$ acceptance. Projection {\em by averaging} of the 4D space to difference axes $(\eta_\Delta,\phi_\Delta)$---a 2D (joint) angular autocorrelation---then discards no information, and correlations are undistorted by the projection~\cite{inverse,flowmeth}.

In contrast, projection to 1D $\phi_\Delta$ or $\phi - \Psi_r$ ($\Psi_r$ -- reaction-plane angle), as in conventional flow analysis, abandons a large amount of information. Different dynamical processes (e.g., flow and nonflow) confused in a 1D azimuth projection may be easily distinguished with a 2D autocorrelation. Examples are shown in Sec.~\ref{azcorr}.


\subsection{Fourier analysis}

In conventional flow analysis a Fourier series represents 1D event-wise azimuth density $\rho(\phi)$~\cite{volzhang,poskvol}. In~\cite{poskvol} the mean value of the azimuth density $\rho_0$ was factored from the Fourier series
\bea \label{eq1}
\rho(\phi) = \rho_0 \left\{  1+2 \sum_{m=1}^\infty v_m \, \cos(m[\phi - \Psi_r]) \right\},
\eea
$\Psi_r$ being the reaction-plane angle. The quantities $v_m$ are then {\em ratios} of true Fourier coefficients. The ratio formulation makes physical interpretation difficult, in part because different dynamical processes contribute to numerator and denominator. 

$v_m$ seems to describe a single-particle distribution, but $\Psi_r$ is not observable. Instead, event-plane (EP) angle $\Psi_m$ is estimated with particles from the collision, and the result is a {\em two-particle correlation analysis} which estimates $\overline{n(n-1)v_m^2}$, $\overline{n\,v_m^2}$ or $\overline{v_m^2}$ depending on algebraic details. Conventional event-plane flow measure $v_2^2\{\text{EP}\}$ approximates $v_2^2\{2\}$ from an exact sinusoid fit to the 1D azimuth autocorrelation.

In contrast, all dynamical processes can be represented by a 2D angular autocorrelation without invoking a physical model. The per-particle angular autocorrelation on $(\eta_\Delta,\phi_\Delta)$ is~\cite{flowmeth}
\bea \label{eq2}
\frac{\Delta \rho}{\sqrt{\rho_{ref}}} \hspace{-.04in} &\equiv& \hspace{-.04in} \frac{\Delta \rho_\text{nf}}{\sqrt{\rho_{ref}}}(\eta_\Delta,\phi_\Delta) \hspace{-.02in} + \hspace{-.02in} 2 \sum_{m=1}^4 \frac{\Delta \rho [m]}{\sqrt{\rho_{ref}}} \cos(m \phi_\Delta),
\eea
where the first term on the RHS describes ``nonflow,'' that is, peaked structures on $\eta_\Delta$ and $\phi_\Delta$, and the second term describes azimuth sinusoid components (multipoles). The $m=3,4$ terms may be measurable but are not essential to understand basic nuclear collision dynamics. We obtain a model-independent per-particle quadrupole component ${\Delta \rho [2]}/{\sqrt{\rho_{ref}}}$ rather than per-pair ``elliptic flow'' measure $v_2$ motivated by specific hydrodynamic expectations~\cite{ollitrault,poskvol}. The relation between the measures
\bea \label{eq3}
\Delta \rho[2] / \sqrt{\rho_{ref}} \equiv  \overline{n(n-1)\langle \cos(2\phi_\Delta) \rangle} / 2\pi \bar n \equiv  \bar n / 2\pi \,\cdot v_2^2,
\eea
defines an {unbiased} version of $\overline{v_2^2}$. Alternative plotting formats incorporating $\Delta \rho[2] / \sqrt{\rho_{ref}}$ and several geometry parameters are considered below. 

\subsection{Azimuth quadrupole $y_z$ dependence}

\begin{figure}[h]
\includegraphics[width=3.3in,height=1.58in]{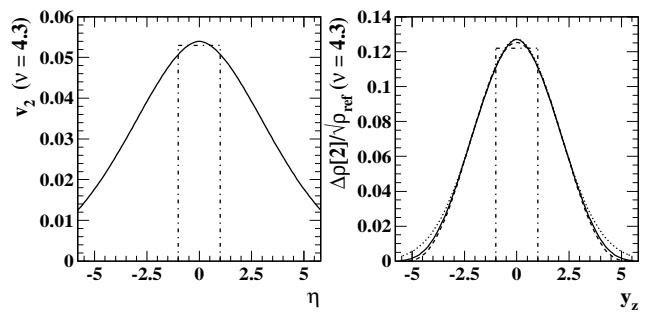}
 \caption{\label{fig5a}
Left panel: Distribution of conventional per-pair flow measure $v_2$ on pseudorapidity $\eta$ for Au-Au collisions at 200 GeV (trend adapted from~\cite{phobflow}). Right panel: The same results converted to per-particle quadrupole measure $\Delta \rho[2] / \sqrt{\rho_{ref}}$ and plotted on pion rapidity $y_z$ assuming $\langle p_t \rangle \sim 0.45$ GeV/c (solid curve). The dotted curve is a gaussian and the dashed curve is a beta distribution, the widths matching the solid curve.  The dash-dot lines denote the STAR TPC acceptance.
 } 
 \end{figure}

The advantage of a per-particle measure is illustrated by the pseudorapidity dependence of $v_2$ over a large $\eta$ interval~\cite{phobflow}. In Fig.~\ref{fig5a} (left panel) $p_t$-integrated data for Au-Au collisions at $\sqrt{s_{NN}} = 200$ GeV are plotted in the conventional format $v_2$ {\em vs} $\eta$. The shape corresponds to reported minimum-bias measurements (the solid curve sketches the data trend), and the peak amplitude agrees with the 40-50\% ($\nu \sim$ 4.3) centrality bin. The rectangles represent the STAR TPC acceptance. Flow appears to have a substantial magnitude even at large pseudorapidity, but errors there are typically consistent with zero.

In Fig.~\ref{fig5a} (right panel) $v_2$ is converted to $\Delta \rho[2] / \sqrt{\rho_{ref}}$ (solid curve) using Eq.~(\ref{eq3}) and the measured $\bar n \rightarrow dn/d\eta$ trend from~\cite{phobmult}. The conversion $\eta \rightarrow y_z$ assumes pions with mean $p_t \sim 0.45$ GeV/c to suggest kinematic limits. The quadrupole component falls to zero within the kinematic limits, and  the statistical errors are approximately uniform on rapidity, another advantage of a per-particle measure. The $\Delta \rho[2] / \sqrt{\rho_{ref}}$ curve is well-described by a gaussian with $\sigma_{y_z} = 2$ (dotted curve), but the gaussian tails extend beyond the kinematic limits. A beta distribution consistent with those limits is plotted as the dashed curve~\cite{lepmini}, with parameters $p = q = 4.8$, r.m.s.~width 1.7 and half-maximum points at $\pm 2.4$.

\subsection{A-A centrality measurement}

Significant improvements in A-A centrality determination were introduced in~\cite{centmeth} based on the approximate power-law form of the minimum-bias distributions on $n_{ch}$ (observed particle multiplicity in an acceptance) and on $n_{part}$ (participant number) and $n_{bin}$ (binary-collision number). Accurate Glauber parameters $n_{part}/2$ and $n_{bin}$ are defined by running integrals on fractional cross section $\sigma/\sigma_0$ from N-N to $b = 0$ A-A collisions.  Accuracy for peripheral collisions is greatly improved by imposing {\em extrapolation constraints} (invoking information from measured p-p collisions). The combination provides centralities accurate to better than 2\% for all geometries. 

Centrality is measured by participant path length $\nu = 2n_{bin} / n_{part}$, the mean number of N-N encounters per participant nucleon pair. Initial-state processes (e.g., parton scattering and fragmentation, minijets) measured {per-participant} should vary linearly with $\nu$ for N-N linear superposition~\cite{axialci,ptscale}.  


\section{A-A eccentricity}

Interpretation of the A-A azimuth quadrupole requires an accurate eccentricity model, but there are major uncertainties about the correct model for nuclear collisions. The relevant model depends on parton $x$. Low-$x$ transverse parton structure could be described by a continuum distribution,  by point-like participant nucleons or by something intermediate. 

Eccentricity relative to the A-A impact parameter is defined as
\bea
\epsilon = \frac{\langle y^2\rangle - \langle x^2 \rangle}{\langle y^2\rangle + \langle x^2 \rangle},
\eea
where $\hat x$ and $\hat z$ define the A-A reaction plane~\cite{volposk}. The definition depends on the weighting function. In the optical Glauber model Woods-Saxon densities represent nuclei A and B. The participant nucleon distribution is determined with thickness functions $T_{A},\,T_{B}$. In a participant Monte Carlo Glauber point-like participant nucleons are used for the weighting. In some cases the reaction plane is determined by the participant nucleons rather than the impact parameter of the parent nuclei~\cite{phobeccent,miller}. The eccentricity then has two components $\pi / 4$ apart (rank-2 tensors on azimuth) combined quadratically as
\bea \label{nonsense}
\epsilon^2_{part} = \frac{\{\langle y^2\rangle - \langle x^2 \rangle\}^2 + \{4\langle x\,y\rangle\}^2}{\{\langle y^2\rangle + \langle x^2 \rangle\}^2}.
\eea

\subsection{Optical Glauber eccentricity}

Some geometry parameters ($b$, $n_{part}$, $\nu$) are best estimated by the participant-nucleon limit (Monte Carlo Glauber model)~\cite{centmeth}. However, at small parton $x$ the partonic azimuth correlation structure is the relevant issue for an eccentricity model. The optical Glauber model describes partons transversely correlated (bounded) only by the nuclear radius, a smooth distribution within that boundary. In the absence of other information that minimally-correlated configuration may be the best description.

\begin{figure}[h]
\includegraphics[width=1.65in,height=1.65in]{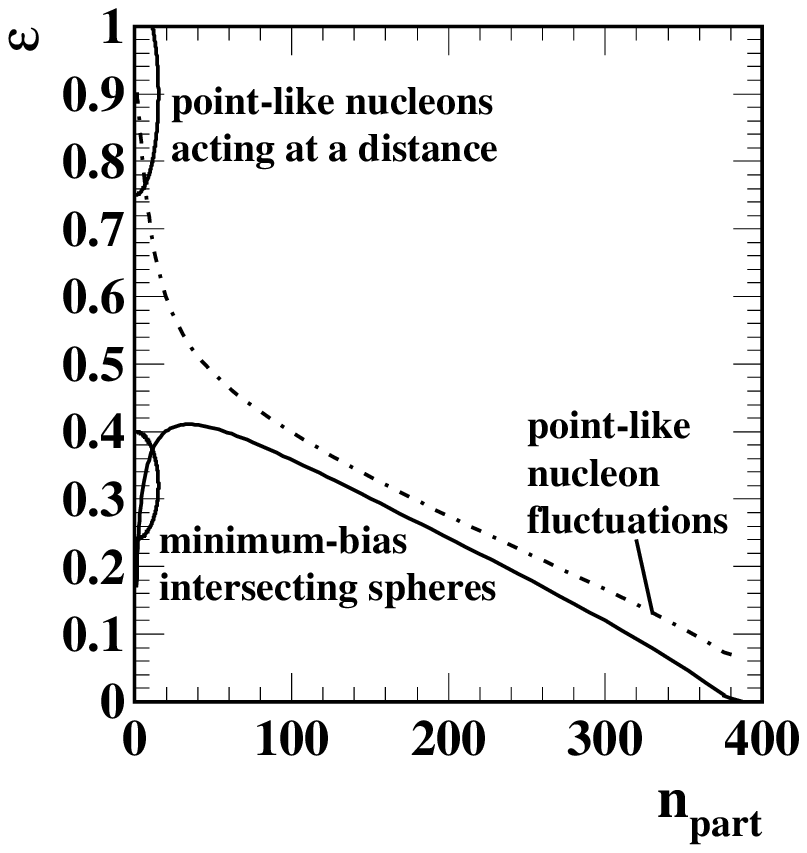}
 \includegraphics[width=1.65in,height=1.65in]{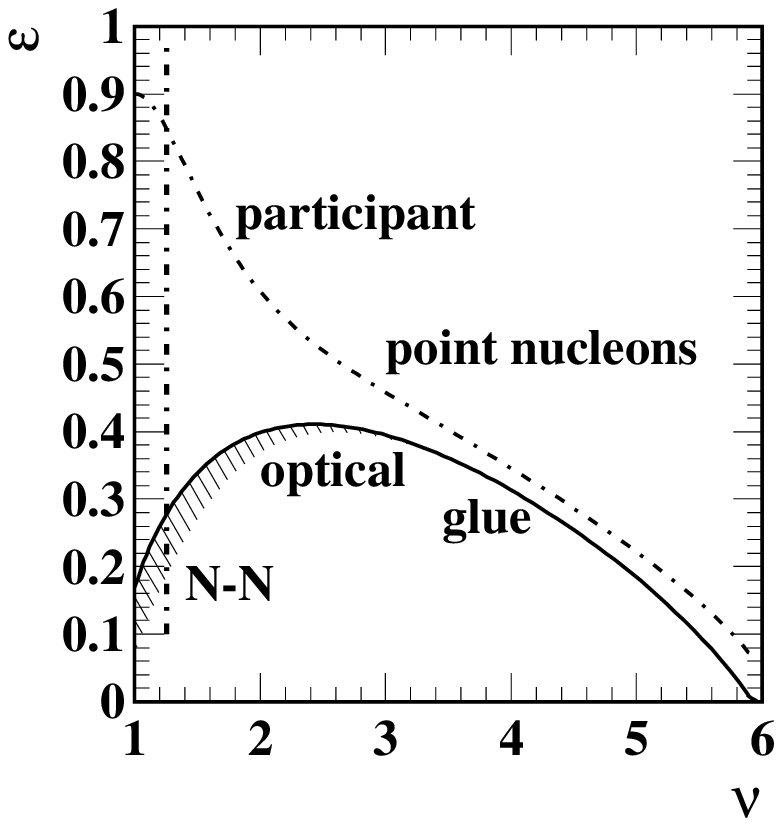}
\caption{\label{fig5}
Left panel: Eccentricity $\epsilon$ {\em vs} participant nucleon number $n_{part}$ modeled by an optical Glauber calculation (solid curve) and by a participant-nucleon Monte Carlo Glauber (dash-dot curve). The latter is a sketch of results in~\cite{manlyeps,voleps}.
Right panel:  The same curves plotted on mean participant path length $\nu$ in number of encountered nucleons. The vertical dash-dot line denotes the mean value of $\nu$ for N-N collisions~\cite{centmeth}.   The hatched region indicates eccentricity uncertainties described in the text.
 } 
 \end{figure}

In Fig.~\ref{fig5}  the solid curves describe an optical Glauber estimate of 17 GeV Au-Au eccentricity with $\sigma_{NN} = 30$ mb described by the power series~\cite{jacoop}
\bea \label{opteccent}
\epsilon(b) &=& -0.047\,(b/b_0)+2.754\, (b/b_0)^2  \\ \nonumber
&-& 4.797\,(b/b_0)^3 +4.852\,(b/b_0)^4-2.492\,(b/b_0)^5.
\eea
Increasing $\sigma_{NN}$ to  40 mb causes a 5\% reduction for central collisions increasing to 13\% for peripheral collisions according to~\cite{jacoop}, small compared to the dramatic difference between optical and participant Monte Carlo Glauber estimates. The hatched region in the right panel indicates the energy dependence of epsilon for peripheral collisions relative to the shape of the 30 mb estimate. We use the optical Glauber Eq.~(\ref{opteccent}) (solid curves) in this paper to interpret quadrupole moments.

\subsection{Participant-nucleon eccentricity}

The dash-dot curves in Fig.~\ref{fig5} are obtained from a participant-nucleon (Monte Carlo) Glauber model and Eq.~(\ref{nonsense})~\cite{phobeccent,manlyeps,voleps}. The event-wise distribution of participant nucleons is used to estimate the reaction plane (which deviates randomly from the A-A impact parameter)~\cite{manlyeps}. The difference between optical and Monte Carlo curves is most dramatic for peripheral and central collisions. 

The large value $\epsilon \rightarrow 1$ for peripheral A-A $\rightarrow$ N-N collisions implies that N-N collisions are {\em on average} rod-like (action at a distance). But nucleons are not point-like objects relative to the average interaction distance. Barring contradictory evidence N-N collisions should be described by the average eccentricity ($\sim 0.25$) of minimum-bias A-A collisions, since the geometry is in either case intersecting spheres.

Nonzero $\epsilon$ for central A-A collisions implies structure resulting from modeling nuclei as distributions of point-like participant nucleons. What justifies that model for parton (gluon) interactions at $x \sim 0.01$? How does that model relate to the color-glass condensate (CGC) as a continuum limit? Since the number of partons is much greater than the number of nucleons at $x \sim 0.01$ there may be substantial central-limit suppression of participant-nucleon correlation structure and fluctuations.


\section{Measured Azimuth Correlations} \label{azcorr}

In~\cite{flowmeth} a simple model for $p_t$-integrated flow centrality dependence was introduced. Flow and minijets (nonflow) inferred from 2D angular autocorrelations on $(\eta_\Delta,\phi_\Delta)$ were compared with 1D projections on azimuth difference variable $\phi_\Delta$. The model was intended to illustrate qualitative features. In this section I pursue the comparisons in more detail and examine novel plotting formats which reveal unexpected simplicity.

\subsection{Fitting 2D angular autocorrelations} \label{geom}

\begin{figure}[h]
  \includegraphics[width=1.65in,height=1.58in]{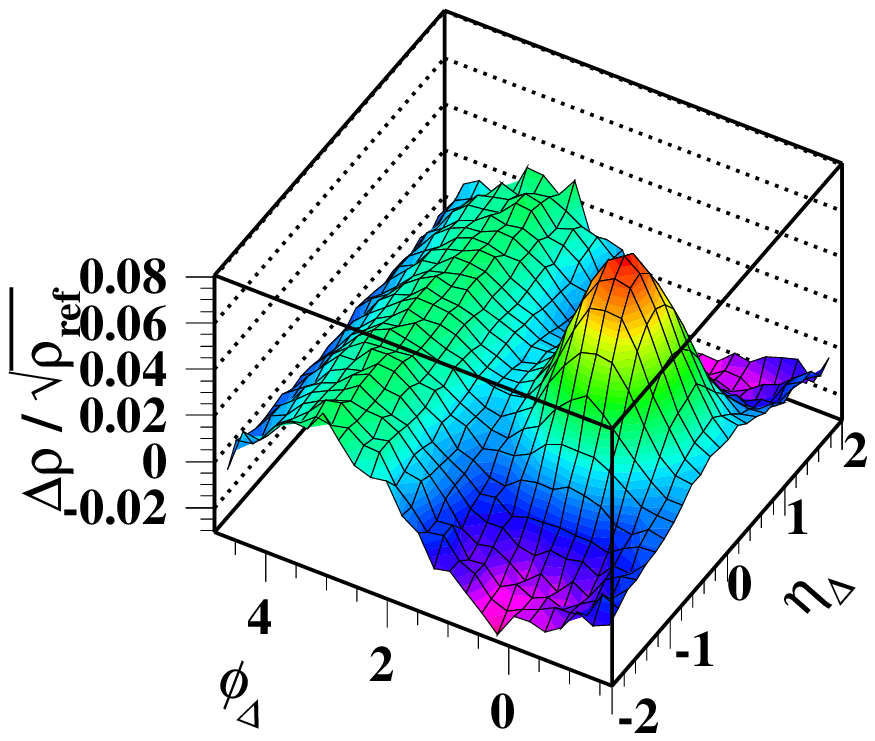}
\includegraphics[width=1.65in,height=1.58in]{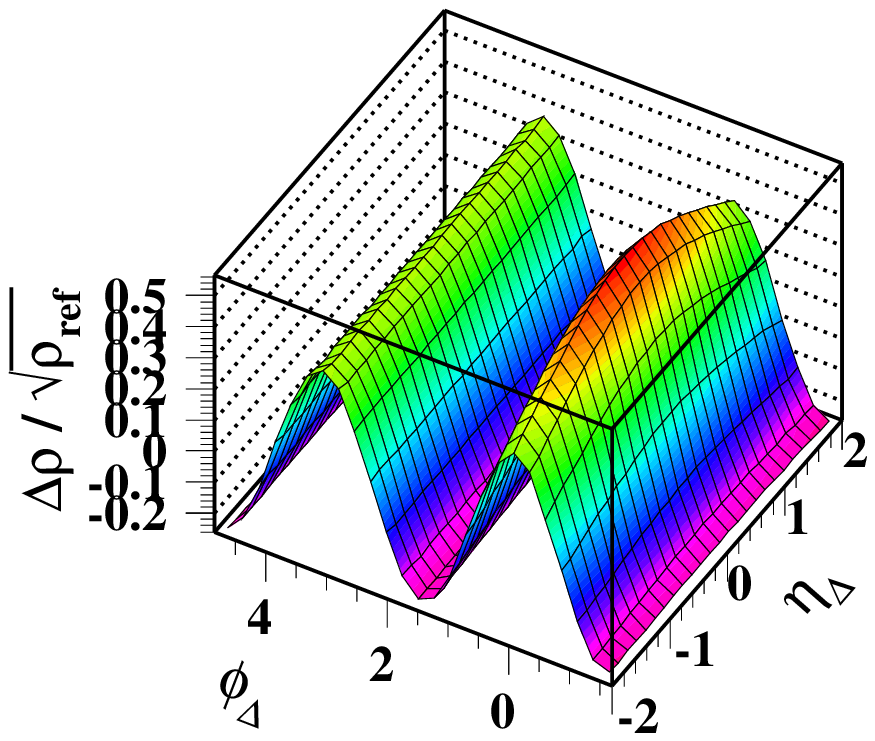}
 \caption{\label{fig2}
Left panel: Simulated 2D angular autocorrelation for 200 GeV Au-Au collisions and 90-100\% centrality ($\sim$N-N collisions). Right panel: autocorrelation for 20-30\% centrality. 
 } 
 \end{figure}

The 2D autocorrelation model for this study is similar to~\cite{flowmeth} with two exceptions. Same-side peak amplitude $0.6\, \nu / 2\pi$ is larger by 15\%, and the quadrupole amplitude model is $\Delta \rho[2] / \sqrt{\rho_{ref}} = 0.0045\, n_{bin}\, \epsilon^2$, as explained in Sec.~\ref{ldl}.  Fig.~\ref{fig2} shows examples for 90-100\% and 20-30\% centrality bins. 

Simulated 2D autocorrelations with statistical noise added (corresponding to about 5M events) for eleven centralities were fitted with a model function consisting of constant offset, azimuth dipole, quadrupole and 2D gaussian same-side peak. 2D autocorrelations were also projected onto $\phi_\Delta$ and fitted with 1D model $\cos(2\phi_\Delta)$ to emulate a $v_2\{2\}$ analysis. Results for 1D and 2D fits are discussed in the next subsection.

\subsection{Per-pair {\em vs} per-particle quadrupole measures}

\begin{figure}[h]
  \includegraphics[width=1.65in,height=1.58in]{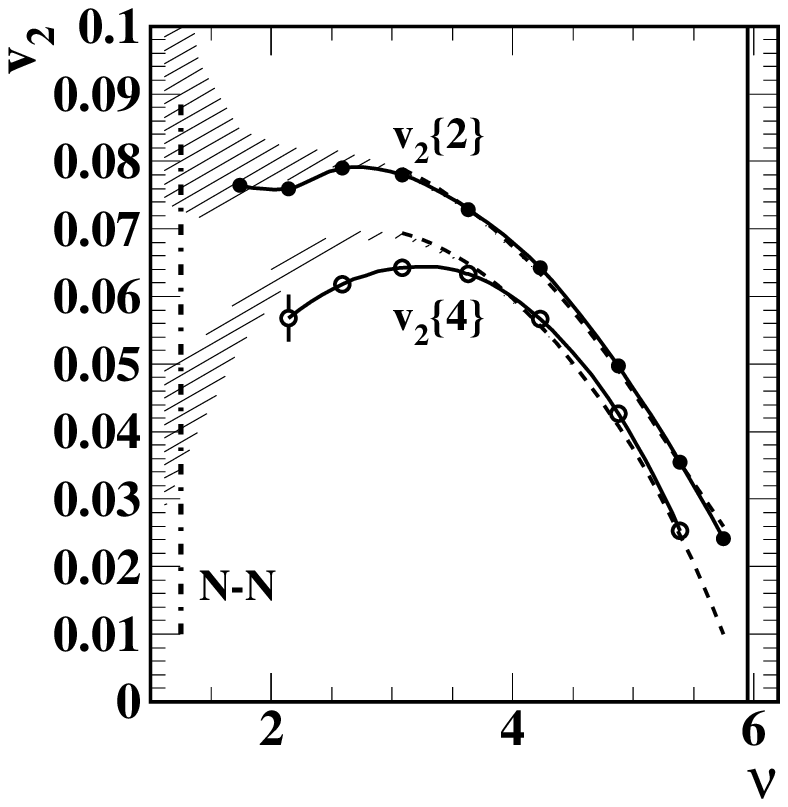}
\includegraphics[width=1.65in,height=1.58in]{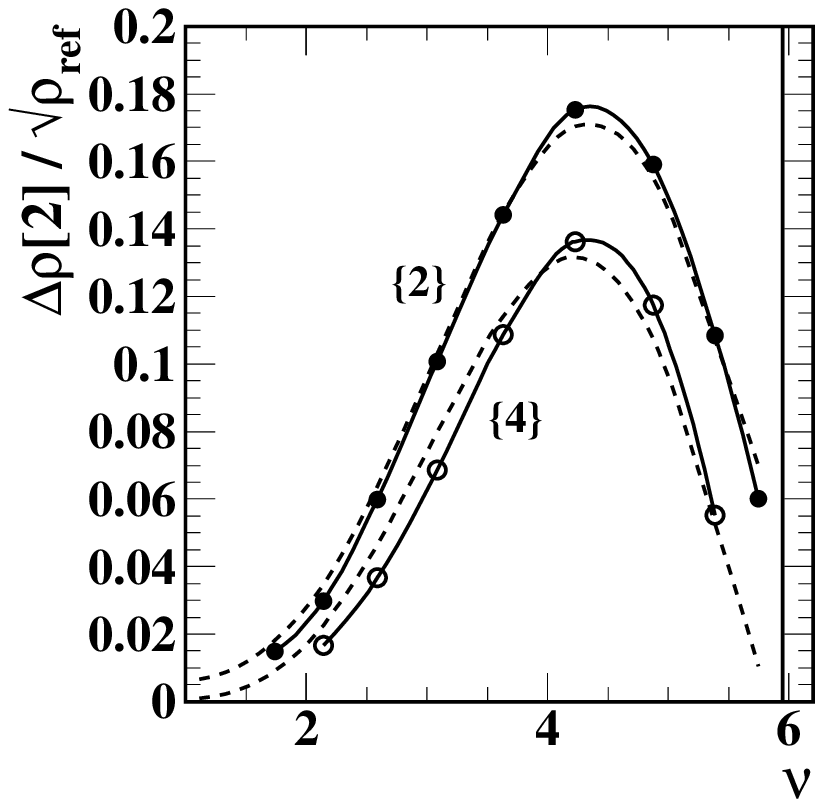}
 \caption{\label{fig1}
Left panel: Conventional per-pair flow measures $v_2\{2\}$ and $v_2\{4\}$, with data from~\cite{2004} (points and solid curves) and corresponding simulations from this work (dashed curves) {\em vs} mean participant path length $\nu$. The vertical dash-dot line denotes N-N collisions~\cite{centmeth}. The hatched regions reflect uncertainties in the definition of $\epsilon$.
Right panel: The same data and curves in terms of per-particle measure $\Delta \rho[2] / \sqrt{\rho_{ref}} $.
 } 
 \end{figure}

Fig.~\ref{fig1} (left panel) shows per-pair $v_2$ {\em vs} participant path length $\nu$, roughly proportional to the fraction of total cross section in the form $1 - \sigma / \sigma_0$~\cite{centmeth}. The points are STAR measurements from~\cite{2002}, and the dashed curves represent 1D and 2D fits to the simulations described above.  The hatched regions indicate the effect of the uncertainty in $\epsilon$ for peripheral collisions shown in Fig.~\ref{fig5} (right panel).

$v_2$ is substantial for peripheral A-A and N-N collisions and falls toward zero for central collisions, typical of per-pair correlation measures which contain an extraneous factor $1/n_{ch}$. The vertical dash-dot line marks the mean position on $\nu$ of N-N collisions~\cite{flowmeth}. $v_2\{2\}$ ``two-particle correlation'' measurements are typically larger than event-plane (EP or ``standard'') measurements~\cite{2004}, although they estimate the same quantity. The $v_2\{4\}$ are four-particle cumulant measurements intended to eliminate ``nonflow'' contributions (minijets)~\cite{borg}. The difference between $v_2\{2\}$ and $v_2\{4\}$ has been interpreted as ``nonflow'' in the past, but more recently has been attributed entirely to $v_2$ fluctuations~\cite{starflucts,phobflucts}. 

In~\cite{flowmeth} the difference between $v_2\{2\}$ and $v_2\{4\}$ was comparable in shape and magnitude to the difference between fits to the 2D angular autocorrelation and its 1D projection (dashed curves). That difference is exactly the $m  =2$ component of the Fourier decomposition of the same-side minijet peak, implying that ``nonflow'' in conventional flow analysis is dominated by crosstalk between minijet correlations and the quadrupole amplitude in a 1D projection on $\phi_\Delta$.

Fig.~\ref{fig1} (right panel) shows per-particle azimuth quadrupole amplitude $\Delta \rho[2] / \sqrt{\rho_{ref}} $ {\em vs} $\nu$. The dashed curves are 1D (upper) and 2D (lower) fits to simulated autocorrelations, transformed from right to left panel {\em via} Eq.~(\ref{eq3}). The $v_2$ data from~\cite{2002} (points) have been transformed from left to right also according to Eq.~(\ref{eq3}). The quadrupole amplitude increases rapidly with centrality to a maximum for mid-central collisions, then follows the trend of the eccentricity toward zero. Although the centrality coverage of measured $v_2\{4\}$ is limited, it is consistent with 2D fits to the simulated autocorrelations which extrapolate to zero at $b = 0$ by construction. The agreement between simulations and data is good despite the simplicity of the model. 

This exercise was intended to demonstrate the effect of minijet correlations on $v_2\{2\}$. The offset between dashed curves is due to the same-side 2D minijet peak, the substantial $m = 2$ component in its 1D azimuth Fourier expansion. The away-side minijet ridge does not contribute to $v_2\{2\}$ bias because it is described by dipole term $\cos(\phi_\Delta)$ and is therefore orthogonal to quadrupole $\cos(2\phi_\Delta)$. The offset between $v_2\{2\}$ and $v_2\{4\}$ is considered further in Secs.~\ref{nonflow} and~\ref{flowflux}.  An unanticipated benefit of the simulation is the agreement of measured $\Delta \rho\{4\} / \sqrt{\rho_{ref}} \equiv \bar n v^2_2\{4\} / 2\pi$ with a simple $n_{bin}\, \epsilon^2$ trend, as described in the next subsection.



\subsection{Binary collisions vs hydrodynamics} \label{ldl}

\begin{figure}[h]
\includegraphics[width=3.3in,height=1.58in]{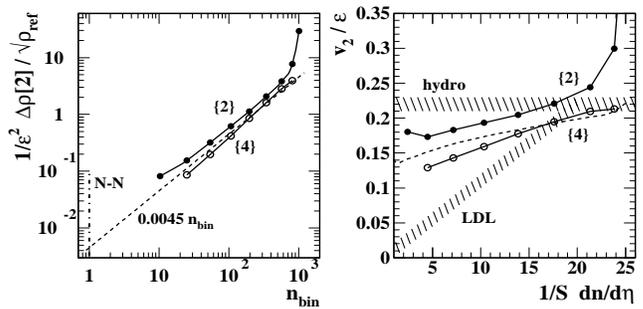}
 \caption{\label{fig3}
Left panel: Per-particle quadrupole measure $\Delta \rho[2] / \sqrt{\rho_{ref}}$ divided by initial-state geometry estimator $\epsilon^2$ (optical Glauber) {\em vs} binary-collisions estimator $n_{bin}$ (Monte Carlo Glauber), showing an approximately linear relation (dashed line) for Au-Au $v_2\{4\}$ data at 200 GeV.
Right panel: Conventional ratio $v_2 / \epsilon$ {\em vs} in-medium collision-number estimator $1/S \, dn_{ch}/ d\eta$. The hatched regions represent the low-density-limit (LDL) expectation (slope arbitrary) and the ideal hydro expectation. The dashed curve is the dashed line in the left panel properly transformed.
 } 
 \end{figure}

In Fig.~\ref{fig3} (left panel) the same $v_2$ data are plotted in the format $1/\epsilon^2\, \Delta \rho[2] / \sqrt{\rho_{ref}}$ {\em vs} $n_{bin}$, the number of binary collisions from a Monte Carlo Glauber simulation. The data derived from $v_2\{4\}$ are well approximated by 
\bea  \label{linear}
1/\epsilon^2\, \Delta \rho[2] / \sqrt{\rho_{ref}} &=& 0.0045\, n_{bin}, 
\eea
suggesting that the azimuth quadrupole may be determined {\em solely by initial-state collision parameters} $(b,\sqrt{s_{NN}},A)$. However, accurate data are lacking, especially for peripheral collisions down to N-N.
The dashed line is the basis for the quadrupole amplitude in the autocorrelation simulations described in Sec.~\ref{geom}. The surprisingly simple linear relation suggests that the physical mechanism of the quadrupole component is the same from N-N to central A-A collisions. It is therefore important to test its validity with accurate data over the broadest possible centrality range and for other collision systems (e.g., lighter A-A and lower energies).

In Fig.~\ref{fig3} (right panel) STAR $v_2$ data are plotted in the conventional format $v_2/\epsilon$ {\em vs} $1/S\, dn/d\eta$, the latter reflecting the low-density limit (LDL) expectation that $v_2 / \epsilon$ increases toward a thermal hydro limit with increasing number of in-medium particle collisions as part of an equilibration process~\cite{heiselberg}. $\epsilon$ from the optical Glauber model is used for this plot so that the two panels of Fig.~\ref{fig3} are consistent. Neither data set follows the hatched LDL trend (its slope is arbitrary). The dashed curve is transformed from the dashed line in the left panel.

The $v_2\{4\}$ data extrapolate to a value for N-N collisions which is more than 60\% of central Au-Au collisions. Should we conclude that there is a strong tendency toward early thermalization and collective expansion in elementary hadronic collisions? We have no experimental evidence for ``saturation'' at a hydro limiting value which might {require} an ideal hydro description. Proximity of data to a hydro prediction at one point does not imply that the hydro model is relevant to collision dynamics. Lack of extended agreement with either theoretical conjecture contrasts with the simple relation in the left panel.


\section{Nonflow {\em vs}\, Flow} \label{nonflow}

Separating elliptic flow from other correlation  sources (nonflow) is an unresolved issue for conventional 1D flow analysis. Several strategies introduced to achieve separation~\cite{poskvol,borg} rely on strong model assumptions about nonflow (e.g., hypothetical trends on A-A centrality and event multiplicity). Methods include higher-cumulant analysis~\cite{borg} and flow-vector analysis~\cite{ollitrault,poskvol,borg}. In~\cite{flowmeth} we introduced azimuth multipole estimation from 2D angular autocorrelations which distinguishes different phenomena without physical model assumptions.

\subsection{Four-particle cumulants}

In~\cite{borg} it was proposed that nonflow bias could be reduced with higher-order cumulants. Higher cumulants (e.g., $v_2\{4\}$) seem to reduce nonflow, but the model-dependent assertion that  $v_2\{4\}$ {\em eliminates} nonflow contamination is questionable if nonflow is minijets as observed. The cumulant argument is based on the assumption that 1) nonflow is dominated by pair correlations, 2) $v_2\{4\}$ is insensitive to pair correlations and 3) true elliptic flow is a universal characteristic of almost all (i.e., $\geq 4$) particles. Assumptions 1) and 3) are not generally true.

Minijets are dominated at smaller $p_t$ by low-$Q^2$ partons fragmenting to hadron pairs and triplets~\cite{fragfunc,minijets}. But the assumption that minijets always contain less than four hadrons, especially at larger $p_t$ and more central collisions, is inconsistent with known fragmentation systematics~\cite{fragfunc} and minijet measurements~\cite{axialci}. The assumption that elliptic flow is always ``carried'' by four or more particles is also unjustified, especially for more peripheral collisions where the quadrupole component may involve fewer than four particles. Thus, substantial bias may survive in $v_2\{4\}$ measurements: positive bias (minijets) for more central collisions and larger $p_t$, negative bias (number of ``flowing'' particles $< 4$) for more peripheral collisions. 


\subsection{Variance of the flow-vector distribution}

Flow, nonflow and flow fluctuations have been studied {\em via} systematic variations of the frequency distribution on ``flow vector'' magnitude $Q_2$ or normalized $q_2 = Q_2 / \sqrt{n}$ (not to be confused with Fourier coefficient ratios $q_m \equiv Q_m / Q_0$ defined in~\cite{flowmeth}). Statistical reference $Q_\text{ref}$ is assumed to be gaussian distributed with variance $\sigma^2_\text{ref} = \bar n$. In the conventional description ``nonflow'' $V_\text{nf}$ also contributes to $Q_2$. Thus, the vector sum and $\tilde Q_{ref}^2$ are
\bea
\vec Q_2 &=& \vec {Q}_{ref} + \vec V_2 + \vec V_\text{nf} \\ \nonumber
\tilde Q_{ref}^2  &=& \tilde Q_2^2 - 2 \tilde Q_2  V_2 \cos(\Psi_2 - \Psi_r) + \tilde V_2^2  \\ \nonumber
 &+& 2  V_2\, \tilde V_\text{nf} \cos(\Psi_\text{nf} - \Psi_r) + \tilde V_\text{nf}^2, 
\eea
where tildes denote event-wise random variables. To simplify I assume that flow $V_2$ is constant in magnitude, and nonflow is not correlated with the reaction plane (neither is true in general). The second sinusoid term is then zero in the mean, $\tilde V_\text{nf}^2 \rightarrow n \tilde g_2$ in conventional notation, and the probability distribution on $\tilde Q_2^2$ is 
\bea \label{qdist}
\frac{dn}{d\tilde Q_2^2} = \frac{1}{\sqrt{2\pi \sigma_n^2}} \exp\left\{  -\frac{ V_2^2 + \tilde Q_2^2}{ \sigma_n^2} \right\} I_0\left\{  \frac{2\tilde Q_2\,  V_2}{ \sigma_n^2} \right\},
\eea
 $\sigma^2_n = \overline{n(1 + g_2)}$ being the variance of $\tilde Q_{2}$ about its mean $V_2$, including nonflow~\cite{ollitrault,poskvol,borg}. The normalized variance difference is~\cite{inverse}
\bea \label{nvardiff}
\Delta \sigma^2_{n/} = \sigma^2_n /  \bar n - 1 = \overline{n\,g_2}/\bar n \equiv g_2 = \bar n \delta_2,
\eea
and $\Delta \sigma^2_{n/} / 2\pi \equiv g_2/2\pi$ (one unit of pseudorapidity) is the nonflow contribution to $\Delta \rho[2] / \sqrt{\rho_{ref}}$ in Eq.~(\ref{eq2}). $\delta_2$ is the nonflow part of $v_2^2\{2\} = \langle \cos(2[\phi_\Delta]) \rangle$ in conventional flow analysis. In a more general treatment $V_2$ may vary event-wise, and part of $\vec V_\text{nf}$ may be correlated with the reaction plane.

We have shown that $g_2$ represents minijet correlations. ``Nonflow'' $g_2$ thus has a nontrivial dependence on centrality and multiplicity reflecting the physics of minijets, a complex subject undergoing intense study~\cite{minijets}. Various properties have been ascribed to $g_2$. In~\cite{2004} $g_2$ is said to increase monotonically with centrality, contrary to previous claims that nonflow is independent of centrality~\cite{2002}.  (In~\cite{2004} the quantity $g'_2 = n_{part}\, g_2 / n_{ch}$ is plotted as $g_2$, and in Eq.~(10) $g''_2 = n_{part} (v_2^2\{2\} - v_2^2\{4\})$ is defined.) However, no simple prescription can exclude minijet contributions from 1D flow analysis. Conventional strategies result in an uncertain mixture of two physical phenomena.

For this study I assume 1) ``nonflow'' $\propto$ minijets and 2) the {\em centrality} dependence is $g_2(\nu) \propto \nu$. The true minijet contribution to $v^2_2\{2\}$ is not strictly proportional to $\nu$; it increases more rapidly because the same-side minijet peak broadens on $\eta_\Delta$ with increasing A-A centrality~\cite{axialci}, thus increasing its 1D projection onto $\phi_\Delta$ relative to the same-side peak amplitude. 
I approximate the difference trend between dashed curves in Fig.~\ref{fig1} (right panel) by 
\bea \label{g2}
g_2 / 2\pi \sim 0.005 \, \nu,
\eea
with $g_2/2\pi$ defined by Eq.~(\ref{nvardiff}) and the comment below it.


\section{Flow Fluctuations and Data} \label{flowflux}

Flow fluctuations have generated considerable recent interest. While the absolute relation of $v_2$ to hydro modeling may be uncertain,  {\em relative fluctuations} could be robust against such uncertainties and might confirm event-wise thermalization ($v_2 \propto \epsilon$). By the same argument, if flow fluctuations are shown to be negligible then the conventional ideal-hydro flow scenario could be threatened. Thus, we examine flow fluctuations in some detail, experimental aspects in this section and theoretical aspects in the next.

\subsection{Interpreting the $v_2\{2\} - v_2\{4\}$ difference}

Recently, differences between $v_2$ methods previously attributed to nonflow have been ascribed instead to $v_2$ fluctuations. From Fig. 2 of \cite{starflucts} we obtain
\bea \label{cumulant}
v^2_2\{2\} &= & \bar v_2^2 +  \sigma^2_{v_2} ~~\text{and}\\ \nonumber
v^2_2\{4\} &\simeq&   \bar v_2^2 -  \sigma^2_{v_2}, ~~\text{therefore}\\ \nonumber
v_2^2\{2\} - v_2^2\{4\} &\simeq& 2\sigma^2_{v_2},
\eea
where $\bar v_2$ is nominally an unbiased mean value. However, in Fig.~\ref{fig1} (right panel) we observe that the difference between $\Delta \rho \{4\} / \sqrt{\rho_{ref}}$ and $\Delta \rho \{2\} / \sqrt{\rho_{ref}}$ [$\bar n\,(v_2^2\{2\} - v_2^2\{4\})/ 2\pi$] corresponds almost exactly to the difference between $\Delta \rho[2] / \sqrt{\rho_{ref}}$ obtained by a fit to a 2D angular autocorrelation (lower dashed curve) and a 1D fit to its projection onto $\phi_\Delta$ (upper dashed curve). Since that difference is clearly identified as minijet structure by its variation on $\eta_\Delta$~\cite{flowmeth}, flow fluctuations inferred from the cumulant relations in Eq.~(\ref{cumulant}) are questionable.

\subsection{Flow fluctuations from the $\tilde Q^2_2$ distribution}

Eq.~(\ref{qdist}) is a conditional probability distribution function or PDF: the distribution on $\tilde Q_2^2$ given a fixed value of $ V_2^2$. If event-wise ``flow'' fluctuates then $\tilde V_2^2$ is also a random variable with its own distribution. The observed $\tilde Q_2^2$ distribution is a folding of the RHS of Eq.~(\ref{qdist}) with
\bea
\frac{dn}{d\tilde V_2^2} &=& \frac{1}{\sqrt{2\pi \sigma_{V_2}^2}}\exp\left\{  -\frac{(V_2 - \bar V_2)^2}{2 \sigma_{V_2}^2} \right\}.
\eea
The PDF includes $\sigma_{V_2}^2$ which broadens the PDF of Eq.~(\ref{qdist}) and which can in principle be obtained from fits to the PDF. However, nonflow also broadens the distribution on $\tilde Q^2_2$, and any single fit to the distribution cannot distinguish the two mechanisms. It is argued that the {\em multiplicity dependence} of the two contributions at fixed centrality is different and can be used to separate flow fluctuations from nonflow. 


To examine the structure of the PDF on $\tilde Q_2^2$ I simplify to 1D, ignoring the out-of-plane component of $\vec Q_2$ to focus on gross systematic trends. Consistent with recent descriptions I rewrite the PDF in terms of $\tilde q_2 \equiv \tilde Q_2 / \sqrt{n}$. Ignoring constant factors the PDF on $\tilde q_2^2$ is given by the convolution integral
\bea
\frac{dn}{d\tilde q_2^2} &\propto& \int d\tilde v_2\, \exp\left\{ -\frac{(\tilde q_2 - \sqrt{n}\, \tilde v_2)^2}{1 + g_2(n)}\right\}  \\ \nonumber
&\times& \exp\left\{ -\frac{(\tilde v_2 - \bar v_2)^2}{2 \sigma^2_{v_2}}\right\}.
\eea
I let $\tilde v_2$ vary about zero mean ($\bar v_2 \rightarrow 0$) and obtain
\bea \label{fold}
\frac{dn}{ d\tilde q_2^2} &\propto& \exp\left\{ -\frac{\tilde q_2^2}{1 + g_2(\nu,n)+ 2n \sigma^2_{v_2}}\right\}.
\eea

It was assumed that any change in the width of Eq.~(\ref{fold}) with {\em random track discard} isolates $v_2$ variance $\sigma^2_{v_2}$, based on the assumption that $g_2$ is approximately independent of multiplicity~\cite{starflucts}. However, systematic studies of minijet correlations show that two-particle correlations, particularly those represented by $g_2$, decrease {\em linearly with random track discard}~\cite{axialci,meanpt}. Since the expected width trend from random discard for $v_2$ fluctuations is $2n \sigma^2_{v_2}$ and that measured for nonflow is $g_2 \propto n$ one cannot distinguish $g_2 / 2n$ from $\sigma^2_{v_2}$ by random discard. ``Flow fluctuations'' inferred from the $\tilde q^2_2$ distribution could as well be minijets, and  minijet correlations may account for all such width variations.

\subsection{Flow fluctuations {\em vs} eccentricity fluctuations} \label{floweccent}

Recent studies of flow fluctuations emphasized the relation of $v_2$ fluctuations and eccentricity fluctuations, and the trend $\sigma^2_{v_2} / v_2^2 \sim \sigma^2_\epsilon / \epsilon^2$ has been claimed~\cite{starflucts,phobflucts}. If  $v_2$ fluctuations are equivalent to eccentricity fluctuations the desired ideal-hydro relation $v_2 \propto \epsilon$ is obtained ``by other means.''  Difference $v_2\{2\} - v_2\{4\}$ has also been attributed to $v_2$ fluctuations. However, the source of  $v_2\{2\} - v_2\{4\}$ was identified by comparing 2D angular autocorrelations and their 1D projections as in Fig.~\ref{fig1}, which demonstrates that the $v_2$ difference interpreted as flow fluctuations is dominated by minijets.

The apparent relation $\sigma^2_{v_2} / v_2^2 \sim \sigma^2_\epsilon / \epsilon^2$ results from two misconceptions: 1) the trend identified as $\sigma^2_{v_2} / v_2^2$ reflects the relation in Fig.~\ref{fig3} (left panel) between the final-state momentum quadrupole component and the initial-state spatial quadrupole moment unrelated to fluctuations, and 2) the quantity identified as $\sigma^2_\epsilon / \epsilon^2$ models the small-$x$ parton distribution as point-like nucleons and NSD N-N collisions as point-like nucleons acting at a distance. 

From Fig.~\ref{fig3} (left panel) we obtain Eq.~(\ref{linear}) $\Delta \rho [2] / \sqrt{\rho_{ref}} \sim 0.005\, n_{bin}\, \epsilon^2$, and from the $v_2\{2\} - v_2\{4\}$ difference in Fig.~\ref{fig1} we obtain  Eq.~(\ref{g2}) $g_2 \sim 2\pi\, 0.005 \nu$. If we misidentify $ g_2/2n \rightarrow $``$\sigma^2_{v_2}$'' and ignore $O(1)$ constant factors we obtain
\bea
\frac{\text{``}\sigma^2_{v_2}\text{''}}{v_2^2} &=& \frac{2\pi \,0.005 \nu}{2n\, v_2^2} \\ \nonumber
&=& \frac{0.005 \nu}{2\Delta \rho[2] / \sqrt{\rho_{ref}}} \\ \nonumber
&=& \frac{\nu}{2\, n_{bin}\, \epsilon^2} \\ \nonumber
&\sim& \frac{1}{n_{part}\, \epsilon^2}.
\eea
But since the participant-weighted eccentricity variance for Poisson statistics is ``$\sigma^2_\epsilon$'' $\equiv \sigma^2_\epsilon\{n_{part}\}\sim 1/n_{part}$ it follows that
\bea
\frac{\text{``}\sigma^2_{v_2}\text{''}}{v_2^2} &\sim& \frac{\text{``}\sigma^2_\epsilon\text{''} }{\epsilon^2}.
\eea
Thus, if the minijet contribution denoted $g_2/2n$ is interpreted as ``$\sigma^2_{v_2}$'' and low-$x$ partons are modeled by point-like nucleons, $v_2$ data and Monte Carlo Glauber seem to support an event-wise $v_2 \propto \epsilon$ connection between initial and final states.

 \begin{figure}[h]
\includegraphics[width=3.3in,height=1.65in]{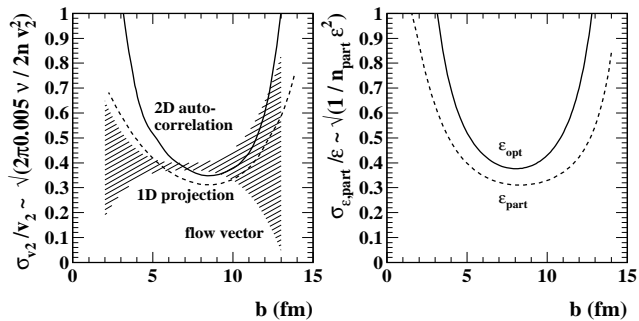}
 \caption{\label{fig4}
Left panel: Relative r.m.s.~$v_2$ fluctuations  {\em vs} A-A impact parameter $b$. The solid curve is derived from minijet trends and $v_2$ obtained from 2D angular autocorrelations. The dashed curve is obtained using minijets and $v_2\{2\}$ from fits to 1D projections. The hatched region represents the flow-vector analysis in~\cite{starflucts}.
Right panel:  An estimate of the relative r.m.s.~eccentricity fluctuations from a participant-nucleon Glauber simulation (see text) divided by optical Glauber eccentricities (solid curve) and participant-nucleon eccentricities (dashed curve). 
 } 
 \end{figure}

In Fig.~\ref{fig4} (left panel ``$\sigma^2_{v_2}$''$/v_2^2 \sim 0.015\, \nu / n\, v_2^2$ (solid curve) results from confusing minijets with $v_2$ fluctuations. The r.m.s.~trend is similar to that in Fig. 2 of~\cite{starflucts}.  The choice of $v_2$ definition in the ratio (solid {\em vs} dashed curves) strongly affects the ratio distribution for peripheral and central collisions. The hatched region sketches $v_2$ fluctuation data and errors from~\cite{starflucts}.
In Fig.~\ref{fig4} (right panel) ``$\sigma^2_{\epsilon}$''$/ \epsilon^2 \sim 1 / [n_{part}\, \epsilon^2]$ (solid curve) results from modeling transverse parton structure by point-like nucleons. Again, the choice among recent definitions of $\epsilon$ strongly affects the r.m.s.~ratio distribution for peripheral and central collisions. For some choices the trends in the two panels are similar, giving the impression that $v_2 \propto \epsilon$~\cite{note}.


\section{Flow fluctuations and theory}

According to some theory expectations flow fluctuations could result from fluctuations during system evolution toward equilibrium as well as from fluctuations in the initial  geometry.  Fluctuations could be generated by in-medium collisions, the onset of turbulence or other aspects of thermalization. Fluctuation measurements might then provide access to equilibration dynamics and medium properties.

\subsection{Flow fluctuations and ideal hydro}

The main evidence for formation of a ``perfect liquid'' at RHIC is successful modeling of $v_2$ centrality and $p_t$ trends by ideal hydrodynamics~\cite{theoryfluct} .  In the ideal hydro model some initial conditions (e.g., spatial azimuth asymmetry) may be {\em smoothly transported} to a manifestation in the final state, implying that fluctuations in the initial geometry could have a direct counterpart in the final state. Recent claims that relative $v_2$ fluctuations are comparable to relative eccentricity fluctuations modeled by the participant  nucleon distribution~\cite{starflucts,phobflucts} seem to validate such expectations.

\subsection{Flow fluctuations and rescattering}

In a non-ideal hydro scenario other aspects of system evolution could contribute to flow fluctuations. Rescattering of a finite number of interacting quasiparticles with mean free path comparable to system size might contribute to $v_2$ variance. If there is significant non-ideal system evolution (rescattering) $v_2$ fluctuations might be sensitive to the {\em Knudson number}.

The Knudson number $Kn = n_{tot} / n_{coll} \sim \lambda / L$ is a measure of internal collisions~\cite{theoryfluct}. The reciprocal Knudson number $Kn^{-1}$ measures the mean number of collisions per DoF during thermalization. Ideal hydro corresponds to $Kn \rightarrow 0$. In~\cite{theoryfluct} a more complete $v_2$ fluctuation description is proposed
\bea
\frac{\sigma^2_{v_2}}{\bar v_2^2} &=& \frac{\sigma^2_{\epsilon}}{\bar \epsilon^2} + \alpha \, Kn.
\eea
Results in~\cite{starflucts,phobflucts} suggest that ${\sigma^2_{v_2}}/{\bar v_2^2} \sim {\sigma^2_{\epsilon}}/{\bar \epsilon^2} \sim 0.15$ for mid-central collisions, and according to~\cite{theoryfluct} $\alpha \sim 1$, implying an upper bound on $Kn \sim 0.03$. 

That experimental upper limit is much smaller than UrQMD predictions, even after invoking artificially large cross sections to speed thermalization~\cite{theoryfluct}. 
The results presented here suggest that the true upper limit on flow fluctuations may be much smaller than 0.15, and $Kn$ may not be meaningful.


\section{Gluonic radiation}


Given the initial and final states of a heavy ion collision we seek the transport mechanisms which intervene. One source of information is propagation of the initial quadrupole moment to the final momentum space. The hydro model provides one description in the form of elliptic flow. Its several parameters ought to have manifestations in the final state. Are they accessible in data, is the hydro model inevitable, are there alternatives?

\subsection{What can hydrodynamics describe?}

The hydrodynamic model  applied to heavy ion collisions is summarized in~\cite{pasi}. 
The hydro sequence parton scattering $\rightarrow$ fast thermalization $\rightarrow$ flow with EoS $\rightarrow$ hadronization provides one scenario for large-scale phase-space transport. However, hydro initial conditions are problematic: pQCD scattering cross sections are too small to thermalize initial-state partons~\cite{molnar}. There appears to be insufficient time with known microscopic processes to unpack the nuclear wave function, equilibrate the results, flow them and reconstitute them into hadrons in the observed final state. In a viscous hydro model the viscosity-to-entropy ratio $\eta / s$ as a model parameter is driven to very small values in attempts to describe data~\cite{rom}. Does that imply a real medium with very small viscosity (perfect liquid) or an inappropriate model?


\subsection{Implications from energy and centrality trends}

In Fig.~\ref{fig3} (left panel) 200 GeV Au-Au $p_t$-integrated $v_2$ data are consistent with 
\bea \label{quad}
 \Delta \rho[2] / \sqrt{\rho_{ref}} &\approx& A\, n_{bin}\, \epsilon^2 \\ \nonumber
\text{or }~~~~ \bar n_{ch} v_2^2/2\pi  &\approx&  \pi A\,  \nu(b)\,\{n_{part}(b)\,\epsilon^2(b)/2\pi\}, 
\eea
where the curly bracket on the RHS represent the per-particle quadrupole moment of the source, and the LHS is the per-particle quadrupole moment of final hadrons. Eq.~(\ref{quad}) suggests that all $p_t$-integrated quadrupole systematics are described by one or two parameters representing the initial A-A system. There is no apparent sensitivity to intervening collision dynamics, no need to invoke a hydrodynamic scenario, equation of state or medium properties such as viscosity. The quadrupole may be completely determined by the initial small-$x$ parton (glue-glue) interaction. 




\subsection{Implications from flow fluctuations}

Reliable separation of the quadrupole component from minijets reveals that flow fluctuations are much smaller than previously claimed. In Sec.~\ref{floweccent} $\sigma^2_{v_2} / v^2_2$ was reduced from mean value 0.15~\cite{starflucts} to {upper limit} 0.03 for $b \sim$ 7 fm. Fluctuations might be much smaller than expected from a participant Monte Carlo simply because there are many more gluons than participant nucleons---a central limit reduction. However, the absence of measurable flow fluctuations may actually hint at the true transport mechanism, a simple relation between the hadronic quadrupole component and initial collision geometry defined by small-$x$ gluonic {\em field} interactions.

\subsection{Generalizing parton scattering}

Parton interactions at large energy scales are modeled in pQCD as point-like interactions. Near the saturation scale, however, QCD interactions should extend over a finite space-time volume---the ``partonic participants'' (interacting fields) may even extend across the nuclear diameter, the model implicit in~\cite{weibel,mrow}.

In a generalization of the pQCD parton-parton vertex to non-pQCD interactions over extended space-time volumes the interaction strength should be the product of a cross section and a {\em relative} current density, a space-time current autocorrelation. Such an interaction would be consistent with Eq.~(\ref{quad}) provided an energy-dependent factor is incorporated.

\subsection{Gluonic multipole radiation}

Given present difficulties with the viscous hydro model, we can ask does a thermalized gluon population evolve according to the hydro scenario? Is a medium-dependent equation of state necessary, or even permitted by data trends? One possibility is that the hydro model is inappropriate. The alternative may be an extended field-field interaction representing a generalization of pQCD. A radially expanding gluonic field might {\em appear} as a flow field, and some hydrodynamic properties (e.g., viscosity) could be ascribed to it.  ``Elliptic flow'' might then be an instance of gluonic quadrupole radiation.

In~\cite{weibel} and references therein interaction of chromo-electromagnetic (CEM) fields leading to Weibel (plasma) instabilities are studied in an attempt to solve the hydro initial conditions problem by replacing the pQCD $2 \rightarrow 2$ partonic collision with extended field-field interactions as a route to fast local thermalization. But the CEM field approach could also supplant the hydro scenario entirely.

The field-field interaction as a transport mechanism solves two major problems for quadrupole phenomenology: 1) the initial-final quadrupole relation Eq.~(\ref{linear}) reveals no information about the dynamical evolution of quasiparticles, and 2) flow fluctuations may be negligible, implying either a large number of DoF or a continuum. Both problems are resolved if continuum field-field interactions provide the transport mechanism. The color filimentation process may provide rapid large-scale transport from longitudinal to transverse phase space, and large azimuthal asymmetries may result~\cite{mrow}. 


Consequences of an initial quadrupole moment have not yet been considered in studies of filimentation and instability. However, once CEM fields are the basis for dynamics multipole radiation is also a possibility. The final-state quadrupole in heavy ion collisions may be a manifestation of gluonic multipole radiation by analogy with Maxwell's equations. Hadron azimuth correlations should reveal an approximately linear superpositions of gluonic multipole sources and corresponding radiated field components. The Weibel instability mechanism may in fact react  to the initial participant eccentricity as a seed for instability growth, resulting in observed large azimuth asymmetries.


 \section{Summary}

I have presented new analysis methods based on 2D angular autocorrelations with emphasis on azimuth correlations and the quadrupole component. I show that fits to 2D autocorrelations achieve model-independent separation of ``elliptic flow'' (quadrupole component) from ``nonflow'', and that the latter is dominated by the same-side minijet peak (jet cone). I have re-examine assumptions about A-A eccentricity and conclude that estimates based on an optical Glauber model better describe low-$x$ parton correlations.

Combining the optical Glauber eccentricity with published elliptic flow data I find a surprisingly simple linear relation to the number of binary collisions. If that relation is confirmed by accurate data over a broader centrality range it is possible that ``elliptic flow'' is dependent only on initial-state collision parameters and is insensitive to hydro parameters or an equation of state.

I also examine recent claims of large elliptic flow fluctuations comparable to participant Glauber eccentricity fluctuations and find that flow-fluctuation measurements are dominated by incorrectly-identified minijet correlations. True $v_2$ fluctuations are much smaller, and present measurements are consistent  with zero.

Given those results and difficulties in accounting for hydrodynamic initial conditions (rapid partonic thermalization), combined with alternative theoretical scenarios invoking QCD field filimentation and instabilities, I conclude that the azimuth quadrupole component may not be a hydrodynamic phenomenon, it may be an example of gluonic multipole radiation.

I appreciate helpful discussions with D.\,T.~Kettler and R.\,L.~Ray.  This work was supported in part by the Office of Science of the U.S. DoE under grant DE-FG03-97ER41020.




\begin{thebibliography}{99}
\bibitem{note} \cite{starflucts} was revised to state that random track discard  cannot differentiate between nonflow (minijets) and flow fluctuations. Previous observations are upper limits, and the analysis is consistent with zero.
\end{thebibliography}
\end{document}